\documentclass[twocolumn]{jpsj3a}

\usepackage{txfonts}
\usepackage{color}

\title{Antiferromagnetism, Superconductivity and Phase Diagram in the
Two-Dimensional Hubbard Model\\ 
-- Off-Diagonal Wave Function Monte Carlo Studies of Hubbard Model III --
}

\author{Takashi Yanagisawa}

\inst{
Electronics and Photonics Research Institute,
National Institute of Advanced Industrial Science and Technology (AIST),
Central 2, 1-1-1 Umezono, Tsukuba, Ibaraki 305-8568, Japan
}

\abst{We investigate the ground-state phase diagram of the
two-dimensional Hubbard model based on the optimization variational Monte Carlo method. 
We use a wave function that is an off-diagonal type given as
$\psi=\exp(-\lambda K)P_G\psi_0$, where $\psi_0$ is a one-particle state,
$P_G$ is the Gutzwiller operator, $K$ is the kinetic operator, and
$\lambda$ is a variational parameter.
The many-body effect plays an important role as an origin of spin
correlation and superconductivity in correlated electron systems.
We examine the competition between the antiferromagnetic state and
superconducting state by varying the Coulomb repulsion $U$, the
band parameter $t'$ and the electron density $n_e$.
We show a phase diagram that includes superconducting and
antiferromagnetic phases and that $t'=0$ is most favorable for
superconductivity. 
}

\begin{document}
\maketitle

\section{Introduction}

The mechanism and properties of high-temperature superconductivity have been studied
vigorously for more than 30 years since the discovery of cuprate
high-temperature superconductors\cite{bed86}.
High-temperature cuprates are typical strongly correlated systems
since the parent materials are Mott insulators when no
carriers are doped.  
It is important to understand the electronic
properties of strongly correlated electron systems
because high-temperature cuprates are typical strongly correlated systems
and the parent materials are Mott insulators when no
carriers are doped.  

The CuO$_2$ plane commonly contained in high-temperature cuprates
consists of oxygen atoms and copper atoms.
The electronic model for this plane is given by the d-p model or
three-band Hubbard
model\cite{eme87,hir89,sca91,ogu94,koi00,yan01,koi01,yan03,koi03,koi06,yan09,
web09,lau11,web14,ave13,ebr16,tam16,web08,hyb89,esk89,mcm90,esk91,has09}.
It appears very difficult to understand the ground-state phase diagram of the d-p model
because of strong correlation between electrons.
The two-dimensional (2D) single-band Hubbard model\cite{hub63,
hub64,gut63,zha97,zha97b,yan96,yan95b,nak97,yam98,yam00,yam11,
har09,yan13a,bul02,yok04,yok06,aim07,miy04,yan08,yan13,yan16}
has been investigated as a simplified model of the d-p model.
The ladder model\cite{noa95,noa97,yam94,koi99,yan95,nak07} has also been studied
in relation to the mechanism of superconductivity in a correlated electron
system.
 
The Hubbard model is one of the fundamental models in 
condensed matter physics.  It was first introduced to understand
the metal-insulator transition\cite{mot74} and is also used to describe 
the magnetic properties
of various compounds\cite{mor85,yos96}. 
By employing the Hubbard model,
we can understand the appearance of inhomogeneous states such as
stripes\cite{tra96,suz98,yamd98,ara99,moc00,wak00,bia96,bia13} and
checkerboard-like density wave states\cite{hof02,wis08,han04,miy09}, 
whose existence was reported for high-temperature cuprates.

Recent studies on the 2D Hubbard model indicate that
a superconducting (SC) phase exists in the ground state\cite{yan16}.
This shows the possibility that the 2D Hubbard model may account for
high-temperature superconductivity.
We show the order parameters of the antiferromagnetic (AF) state
and SC state as a function of the interaction parameter $U$ in Fig. 1.
The result shows that high-temperature superconductivity may occur
in the strongly correlated region of the Hubbard model where the
interaction $U$ is greater than the bandwidth.

A variational Monte Carlo method is a useful tool to investigate the
electronic properties of strongly correlated electron systems when
we calculate the expectation values numerically\cite{nak97,yam98,yam00,yam11,har09,yan13a}.
In general, a variational wave function is improved by introducing
new variational parameters to control the electron
correlation.  In our method the wave functions are
optimized by multiplying an initial function
by $\exp(-S)$-type
operators\cite{yan16,yan98}, where $S$ is a suitable correlation operator.
The Gutzwiller function is also written in this form.
An optimization process is performed in a systematic way by
multiplying by the exponential-type operators repeatedly\cite{yan98}.
The ground-state energy is indeed lowered considerably by using
this type of wave function\cite{yan16}.

\begin{figure}
\centering
\includegraphics[width=6.5cm]{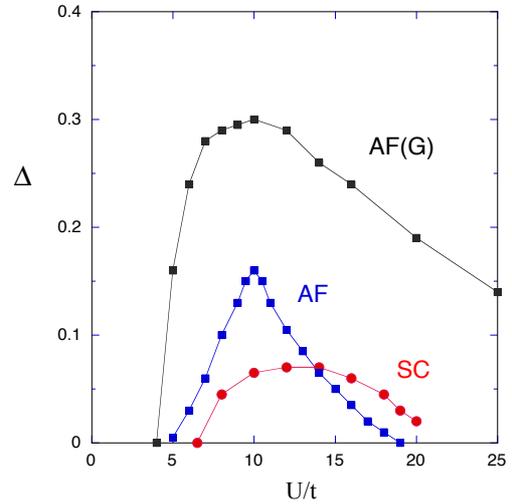}
\caption{(Color online)
AF and SC order parameters as a function of
$U/t$ on a $10\times 10$ lattice with the periodic
boundary condition in one direction and the antiperiodic one in the
other direction\cite{yan16}.  
In Ref. 39, $\Delta$ was shown as a function of $U$ in the
range $0<U<20$.  Here we include the range
$20<U<25$.  AF(G) indicates the result for the Gutzwiller wave function. 
Some data have been updated.
}
\label{fig1}
\end{figure}

In this paper we investigate the stability of the antiferromagnetically
ordered state and show the phase diagram of the ground state of
the 2D Hubbard model.  In the strongly correlated region, the AF
correlation is suppressed and the SC correlation is enhanced.
Near the boundary of the AF region, a large spin fluctuation is
induced, which is considered to give rise to high-temperature superconductivity. 
The paper is organized as follows.  In Sect. 2, we present the
model Hamiltonian and wave functions that we use in the optimization
variational Monte Carlo method.  In Sect. 3, we examine the
wave function with the AF order parameter to show the region where the
AF state is stabilized.
In Sect. 4, we discuss the phase separation that may occur near
half-filling.  In Sect. 5, we investigate the SC state
and show the phase diagram.
We give a summary in the final section.

\section{Optimization Variational Monte Carlo Method}

\subsection{Hamiltonian}

The Hubbard model is written as
\begin{equation}
H= \sum_{ij\sigma}t_{ij}c^{\dag}_{i\sigma}c_{j\sigma}
+U\sum_in_{i\uparrow}n_{i\downarrow},
\end{equation}
where $t_{ij}$ indicates the transfer integral and $U$ is the strength
of the on-site Coulomb interaction.
We set $t_{ij}=-t$ when $i$ and $j$ are nearest-neighbor pairs $\langle ij\rangle$
and $t_{ij}=-t'$ when $i$ and $j$ are next-nearest-neighbor pairs.
We consider this model in two dimensions, and $N$ and $N_e$ denote
the number of lattice sites and the number of electrons, respectively.
The energy unit is given by $t$.

\subsection{Off-diagonal wave function}

In a variational Monte Carlo method, we employ a wave function that is
suitable for the system being considered and evaluate the expectation
values by using a Monte Carlo procedure.  To take into account the
correlation between electrons, we start from the Gutzwiller wave
function given by
\begin{equation}
\psi_G = P_G\psi_0,
\end{equation}
where $P_G$ is the Gutzwiller operator 
$P_G= \prod_j(1-(1-g)n_{j\uparrow}n_{j\downarrow})$, where $g$ is the
variational parameter in the range of $0\le g\le 1$.  $\psi_0$
indicates a trial one-particle state.

Because the Gutzwiller function is very simple and is not enough to
take account of electron correlation, we should improve the wave
function.  There are several methods to improve the wave function.
One method is to multiply the Gutzwiller function by an 
exponential-type operator.  The wave function is written 
as\cite{yan16,yan98,ots92,yan99,eic07,bae09,bae11,yan14}
\begin{equation}
\psi_{\lambda}= \exp(-\lambda K)\psi_G,
\label{wf1}
\end{equation}
where $K$ is the kinetic part of the Hamiltonian and $\lambda$ is
a real variational operator\cite{yan13a,yan98,yan99}.
The expectation values are calculated by using the auxiliary field
method\cite{yan98,yan07}.
The other method is to introduce
a Jastrow-type operator\cite{yok04}.  We control the nearest-neighbor
correlation by multiplying by the operator
\begin{equation}
P_{Jdh} = \prod_j\left( 1-(1-\eta)\prod_{\tau}\Big[
d_j(1-e_{j+\tau})+e_j(1-d_{j+\tau})\Big] \right),
\end{equation}
where $d_j$ is the operator for the doubly occupied site given as
$d_j=n_{j\uparrow}n_{j\downarrow}$ and $e_j$ is that for
the empty site given by $e_j=(1-n_{j\uparrow})(1-n_{j\downarrow})$.
$\eta$ is the variational parameter in the range
$0\le \eta\le 1$.  With this operator, we can include the 
doublon-holon correlation:
\begin{equation}
\psi_{\eta}=P_{Jdh}\psi_G.
\end{equation}
It is possible to generalize the Jastrow operator to consider
long-range electron correlation by introducing new variational
parameters\cite{mis14,yok17}.

In this paper we use the wave function of the exponential type in
eq. (\ref{wf1}).  We call this type of wave function the off-diagonal
wave function since the off-diagonal correlation in the site
representation is taken into account in this wave function.
We believe that it is more important to consider off-diagonal 
electron correlation than the diagonal electron correlation.
In fact, the energy is further lowered when we employ the
off-diagonal wave function\cite{yan16}.

\subsection{Antiferromagnetic state}

 The AF one-particle state $\psi_{AF}$ is given by the eigenfunction
of the AF trial Hamiltonian
\begin{equation}
H_{AF}= \sum_{ij\sigma}t_{ij}c^{\dag}_{i\sigma}c_{j\sigma}
-\Delta_{AF}\sum_{i\sigma}(-1)^{x_i+y_i}\sigma n_{i\sigma},
\end{equation}
where $\Delta_{AF}$ is the AF order parameter and $(x_i,y_i)$
represents the coordinates of site $i$.
The wave function is written as
\begin{equation}
\psi_{\lambda,AF}= \exp(-\lambda K)P_G\psi_{AF}.
\end{equation}
In general, the AF state is very stable in the Hubbard model near
half-filling.  Thus,
it is important to control the AF magnetic order so that the SC
state is stabilized and realized.

The stability of the AF state depends mainly on the electron density
$n_e$, the interaction strength $U$,
the transfer integral $t'$, and long-range transfers in the 
single-band Hubbard model. 
The AF correlation is induced as $U$ increases from zero in the weakly
correlated region and is maximized when $U$ is of the order of the
bandwidth, say at $U=U_c$, when carriers are doped.  
When $U$ becomes larger than $U_c$,
the AF correlation starts to decrease.  In the region where $U$ is
extremely large, the AF correlation is suppressed to a small
value by the large fluctuation.  This is shown in Fig. 1.  
This is a crossover
between the weakly correlated region and strongly correlated region.

\subsection{Correlated superconducting state}

The SC state is represented by the BCS wave function
\begin{equation}
\psi_{BCS}= \prod_k(u_k+v_kc^{\dag}_{k\uparrow}c^{\dag}_{-k\downarrow})
|0\rangle,
\end{equation}
with coefficients $u_k$ and $v_k$ that appear in the ratio
$u_k/v_k=\Delta_k/(\xi_k+\sqrt{\xi_k^2+\Delta_k^2})$, where
$\Delta_k$ is the gap function with ${\bf k}$ dependence and
$\xi_k=\epsilon_k-\mu$
is the dispersion relation of conduction electrons.
We assume $d$-wave symmetry for $\Delta_k$:
$\Delta_k= \Delta_{SC}(\cos k_x-\cos k_y)$.
The Gutzwiller BCS state is formulated as
\begin{equation}
\psi_{G-BCS}=P_{N_e}P_G\psi_{BCS},
\end{equation}
where $P_{N_e}$ indicates the operator used to extract the state with $N_e$
electrons.  In this wave function the electron number is fixed and
thus the chemical potential in $\xi_k$ is regarded as a variational
parameter.
In the formulation of $\psi_{\lambda}$, we use the BCS wave function
without fixing the total electron number, namely, without the operator
$P_{N_e}$.  The chemical potential $\mu$ in $\xi_k$ is not a variational
parameter and is used to adjust the total electron number.
The wave function is given as
\begin{equation}
\psi_{\lambda}= e^{-\lambda K}P_G\psi_{BCS}.
\end{equation}
We perform the electron-hole transformation for down-spin electrons:
\begin{equation}
d_k= c^{\dag}_{-k\downarrow},~~~ d^{\dag}_k= c_{-k\downarrow},
\end{equation}
and not for up-spin electrons: $c_k= c_{k\uparrow}$.
The electron pair operator $c^{\dag}_{k\uparrow}c^{\dag}_{-k\downarrow}$
denotes the hybridization operator $c^{\dag}_kd_k$ in this formulation.

\section{Antiferromagnetic phase}

Since the SC state competes with the AF state,
it is important to clarify the region of the AF state in the parameter
space.  We use $U$ and $t'$ to control the strength of the AF correlation.
The Coulomb interaction $U$ is important since the magnitude of the AF
magnetism can be controlled by changing $U$.
The transfer integral $t'$ is also important and shows nontrivial effect on
the stability of the AF magnetic order.
One may expect that the AF region will be small when including $t'$
in the model.  This is not, however, true.
As $|t'|$ increases, the AF correlation increases, where
we assume negative $t'$ in this paper.
From the viewpoint of competition between superconductivity and
AF ordering, $t'=0$ is most favorable for
superconductivity.

We show the condensation energy $\Delta E_{AF}$ due to the AF
magnetic order as a function of $1-n_e$ in Fig. 2 for $t'=0$.
When $U$ is as large as $U/t\ge 14$, the AF region exists up to 10$\%$ 
doping.
When $t'=-0.2t$, the AF region expands up to about
20$\%$ doping where $1-n_e\sim 0.2$ even for large $U$.
We show this in Fig. 3.
The AF region becomes larger as $|t'|$ increases.
Figure 4 shows $\Delta E_{AF}$ as a function of $x$ for
$t'=0$, $-0.1$, and $-0.2$, where we use $U/t=18$.
We show the AF region on the $x-t'$ plane in Fig. 5.
The AF state dominates near half-filling and is stabilized as $|t'|$
increases.  The $d$-wave SC state exists near the
boundary in Fig. 5.

\begin{figure}
\centering
\includegraphics[width=8.0cm]{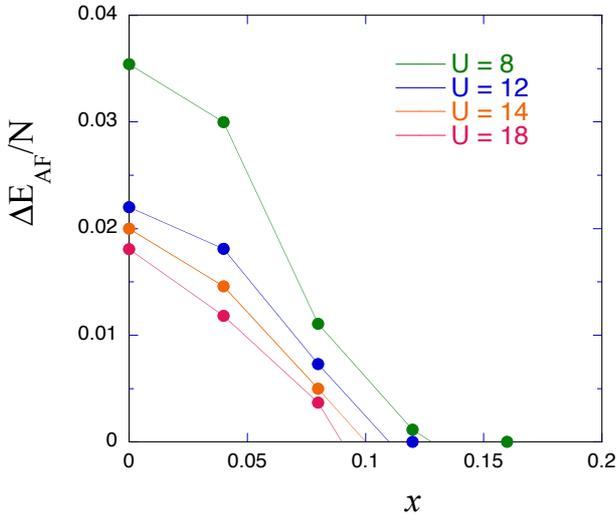}
\caption{(Color online)
The condensation energy of the AF state $\Delta E_{AF}$ as a function 
of the hole density
$x\equiv 1-n_e$ on a $10\times 10$ lattice for $t'=0$.
We put $U/t=8, 12, 14$, and 18.
}
\end{figure}

\begin{figure}
\centering
\includegraphics[width=8.0cm]{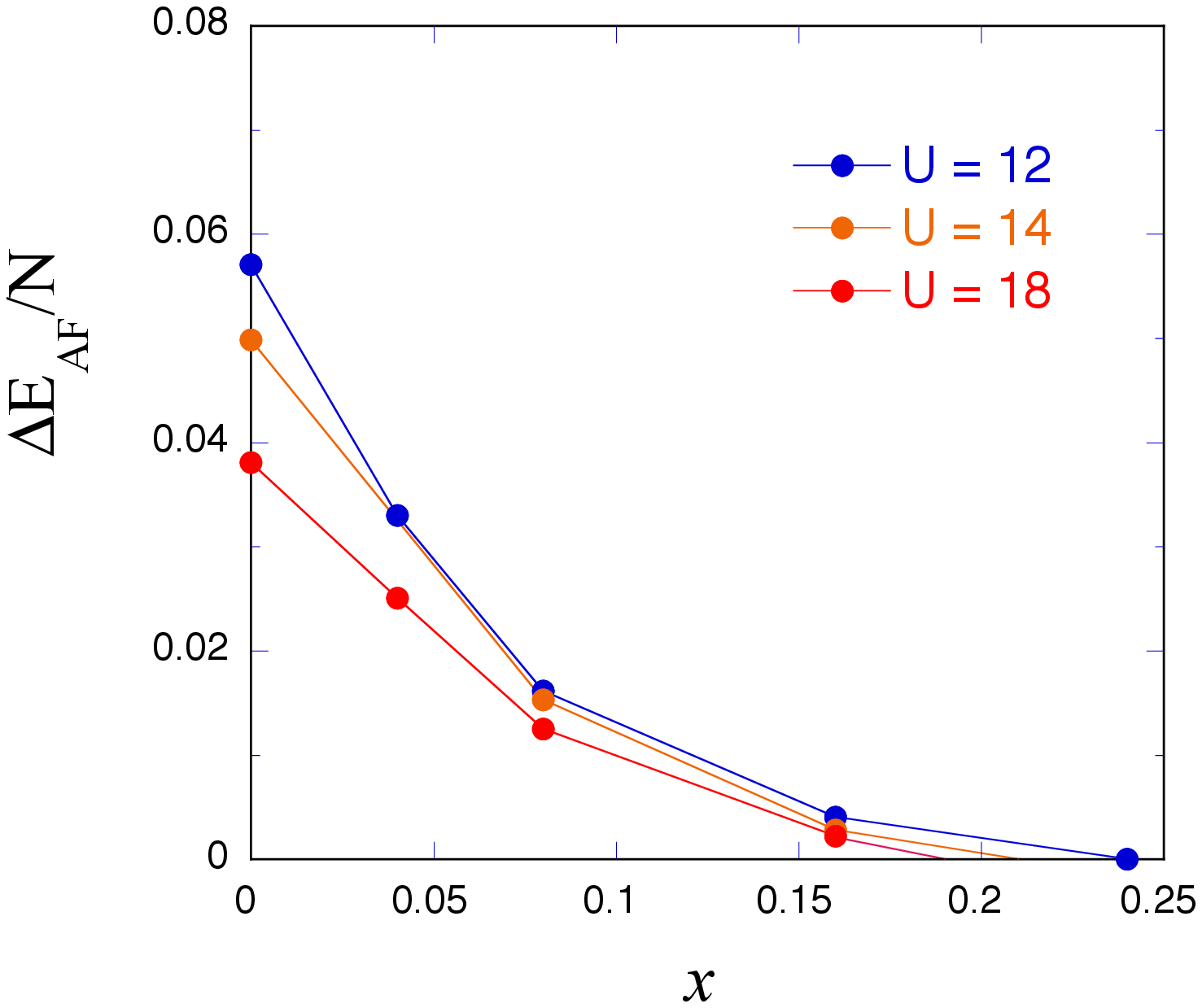}
\caption{(Color online)
The condensation energy of the AF state $\Delta E_{AF}$ as a function 
of the hole density
$x=1-n_e$ on a $10\times 10$ lattice for $t'=-0.2t$.
We put $U/t=12, 14$, and 18.
}
\end{figure}

\begin{figure}
\centering
\includegraphics[width=8.0cm]{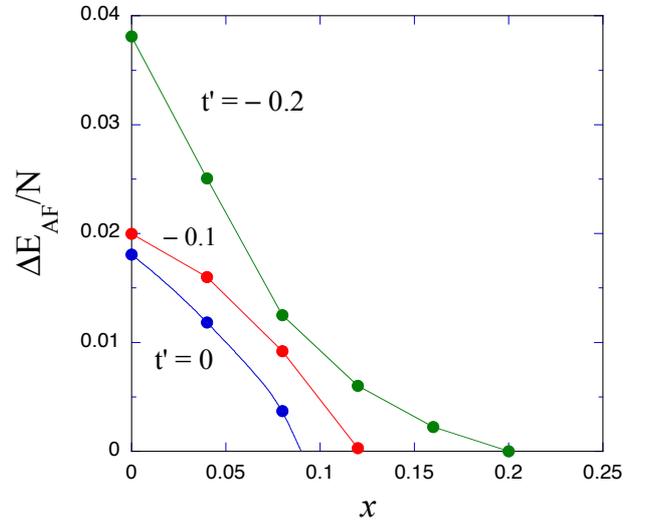}
\caption{(Color online)
The condensation energy of the AF state $\Delta E_{AF}$ as a function 
of the hole density
$x= 1-n_e$ on a $10\times 10$ lattice for $U/t=18$.
From the top we set $t'/t=-0.2, -0.1$, and 0.
}
\end{figure}

\begin{figure}
\centering
\includegraphics[width=8.0cm]{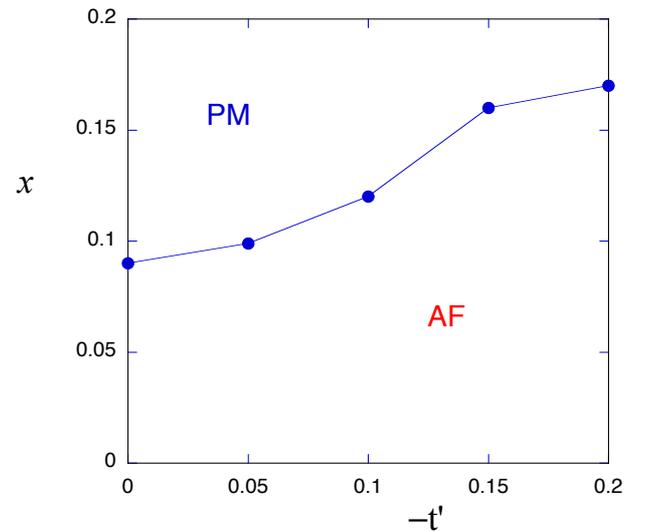}
\caption{(Color online)
Antiferromagnetic and paramagnetic states in the plane
of hole density $x$ and $t'$ for $U=18$.
The energy unit is given by $t$. 
}
\end{figure}

\section{Phase Separation}

We discuss the phase separation in the 2D Hubbard model in this section.
The existence of phase separation near half-filling
in the 2D Hubbard model has been discussed\cite{and93,cos98,mac06}.
This is a subject concerning the AF correlation and charge distribution
in the case of small doped carriers.
When there is a strong AF correlation between neighboring 
electrons, there may be a tendency that doped holes form clusters due
to an effective attractive interaction between electrons.
This means the possibility of phase separation with clusters of holes,
depending on the AF correlation, attractive interaction, and
kinetic energy gain.
This is similar to the instability in the t-J model\cite{sat18}.

We examine an instability toward phase separation by evaluating the
charge susceptibility
\begin{equation}
\frac{1}{\chi_c}= \frac{\partial ^2E(N_e)}{\partial N_e^2}
= \frac{E(N_e+\delta N_e)+E(N_e-\delta N_e)-2E(N_e)}{(\delta N_e)^2},
\end{equation}
where $E(N_e)$ is the ground-state energy and $N_e$ is the number of
electrons. 
This is proportional to the second derivative of the energy $E(N_e)$ with
respect to the electron number.
The negative sign of $\chi_c$ indicates an instability toward the
phase separation.  This instability is very subtle.
Once the phase separation occurs, the ground state becomes an insulating
state.

We show the energy as a function of the doping rate $x$ for $t'=0$
in Fig. 6.  The curve of the energy is usually convex downward,
that is, $\chi_c>0$, but the sign
of $\chi_c$ changes in the region near half-filling.
We show
\begin{equation}
\delta^2E(N_e)\equiv E(N_e+\delta N_e)-2E(N_e)+E(N_e-\delta N_e)
\end{equation}
for $N_e=2$ in Fig. 7.  This indicates that there is an instability
toward the phase separation when $x<0.06$ for $U/t=18$ and $t'=0$.
This is similar for $U/t=14$ and 12.
When $t'$ is nonzero and negative, the instability toward the phase separation
occurs for a smaller doping rate.  We show $d^2E$ for $t'=-0.2$
in Fig. 8.  In this case, $\chi_c>0$ for at least $x>0.06$.
The phase separation area decreases for $t'<0$.

Figure 8 indicates that $d^2E$ shows a depression near $x=0.12$
for $t'<0$.  This suggests
the existence of strong charge fluctuation.
We expect that this shows an instability toward some charge-ordered
state such as the striped state.

In our optimized wave function, the instability toward the phase separation
is limited to the range $x\equiv 1-n_e \le 0.06$ for $t'=0$ and
the region of phase separation becomes small for negative $t'$. 
We also mention that there is a possibility that the phase separation
area will decrease
as the wave function is optimized further by multiplying by
operators $P_G$ and $\exp(-\lambda' K)$.

\begin{figure}
\centering
\includegraphics[width=8.0cm]{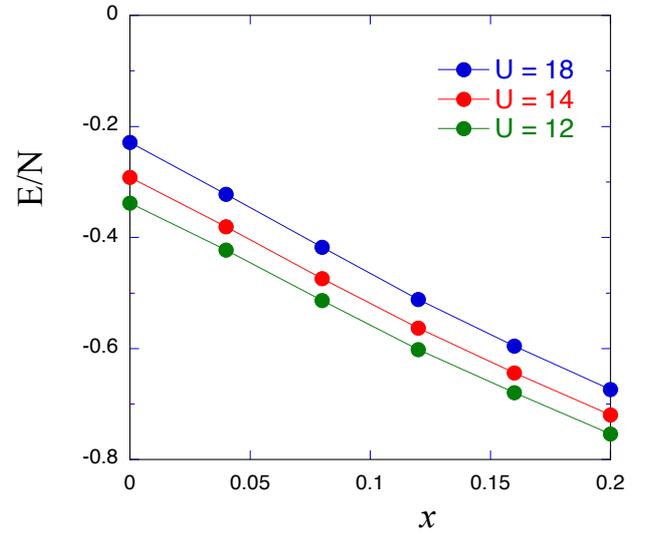}
\caption{(Color online)
The ground-state energy as a function of the hole density $x$
for $U=18t$, $14t$, and $12t$ on $10\times 10$ lattice where
we set $t'=0$.
}
\end{figure}

\begin{figure}
\centering
\includegraphics[width=8.0cm]{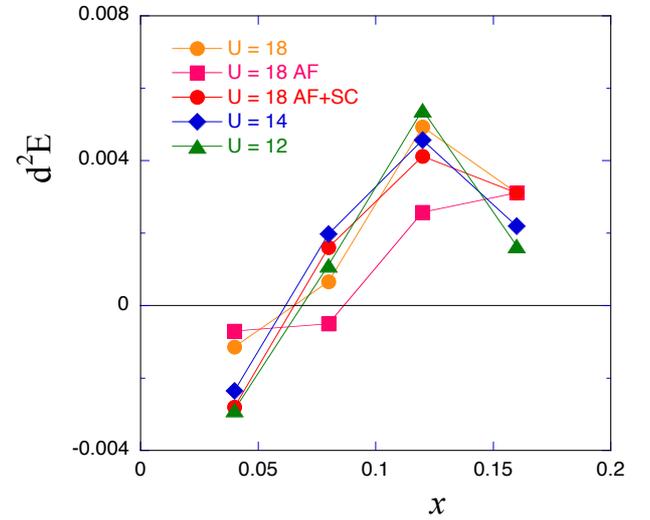}
\caption{(Color online)
The second derivative $d^2E$ as a function of the hole density $x$
for $t'=0$.  AF means the result for the AF state,
where we have introduced the AF order parameter.
The energy unit is given by $t$. 
}
\end{figure}

\begin{figure}
\centering
\includegraphics[width=8.0cm]{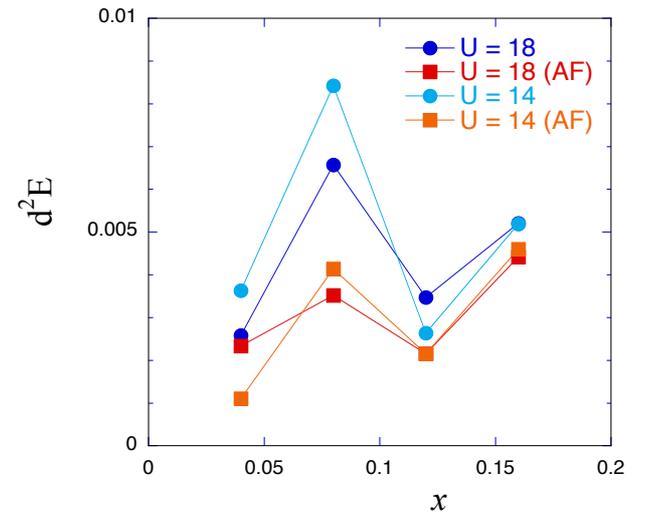}
\caption{(Color online)
The second derivative $d^2E$ as a function of the hole density $x$
for $t'=-0.2$.
The energy unit is given by $t$. 
}
\end{figure}

\section{Superconducting phase}

Let us now discuss the SC state.
We evaluate the SC condensation energy defined by
\begin{equation}
\Delta E= E(\Delta_{SC}=0)-E(\Delta_{SC,{\rm opt}}),
\end{equation}
where $\Delta_{SC}$ is the SC order parameter and $\Delta_{SC,{\rm opt}}$
is its optimized value.
$\Delta E$ is the energy lowering
due to the inclusion of the SC order parameter.
We show the $U$-dependence of $\Delta_{SC,{\rm opt}}$ in Fig. 9 on a $10\times 10$
lattice for $N_e=88$ and $t'=0$, where the upper curve is for the
BCS-Gutzwiller function and the lower one is for the $\psi_{\lambda}$
function.  Both curves in Fig. 9 have a maximum at $U/t\approx 12-14$.
This shows that high-temperature superconductivity is possible in the
strongly correlated region.

We discuss here the coexistence of SC and AF phases in the range of
$0.06<x<0.09$.  The wave function is written in the form
\begin{equation}
\psi_{\lambda}= e^{-\lambda K}P_G\psi_{BCS-AF}.
\end{equation}
The BCS wave function with both the SC order parameter $\Delta_{SC}$ and
the AF order parameter $\Delta_{AF}$ is formulated by solving the
Bogoliubov equation\cite{miy04,yan09}.
We show the ground-state energy $E$ as a function of the electron number $N_e$
near $x\sim 0.08$ for $U/t=18$ in Fig. 10 for $\Delta_{SC}= 0.005$, 0.01, 0.02, and 0.03,
where the chemical potential $\mu$ in the BCS wave function is changed to
adjust the number of electrons. 
The energy $E$ is lowered slightly by introducing $\Delta_{SC}$, by about
$\Delta E/N \sim 0.005t$ per site at $\Delta_{SC}\sim 0.01t$. 
Here we used the parameters $g=0.0018$, $\lambda=0.055$, and $\Delta_{AF}=0.19t$.

The condensation energy per site as a function of the hole density (doping rate) $x$ is 
shown in Fig. 11 for $U/t=18$ and
$t'=0$ on a $10\times 10$ lattice. 
There is an instability toward the phase separation for $x\le 0.06$.
Thus, the AF state for $x\le 0.06$ is an AF
insulator.
There is a coexistent metallic phase of superconductivity and
antiferromagnetism when the doping rate is $0.06\le x<0.09$.
The SC condensation energy at $x=0.08$ is that for the coexistent state.
In the range $0.09<x$, the pure $d$-wave SC state exists.
We have also presented the result obtained by using the level-4 function
with the AF order parameter in Fig. 11:
\begin{equation}
\psi^{(4)}= e^{-\lambda' K}P_G(g')e^{-\lambda K}P_G(g)\psi_{AF},
\end{equation}
where $g$, $g'$, $\lambda$, and $\lambda'$ are variational parameters.
The condensation energy of the AF state for $\psi^{(4)}$ is less
than that for $\psi_{\lambda}=\psi^{(2)}$. 

With the inclusion of $t'$, the phase separation region will decrease,
and at the same time the area of the AF metallic state will increase. 
The phase diagram is dependent on $t'$.

\begin{figure}
\centering
\includegraphics[width=7.9cm]{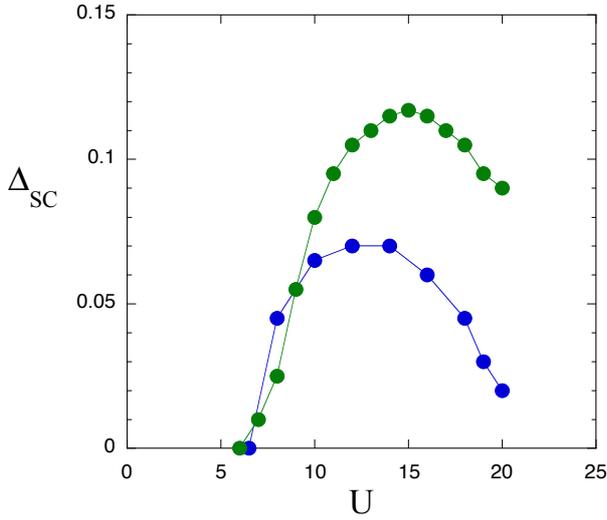}
\caption{(Color online)
The SC order parameter as a function of $U$
on a $10\times 10$ lattice with $t'=0$.
We used the BCS-Gutzwiller function for the upper curve and
$\psi_{\lambda}$ for the lower curve.
}
\end{figure}

\begin{figure}
\centering
\includegraphics[width=7.9cm]{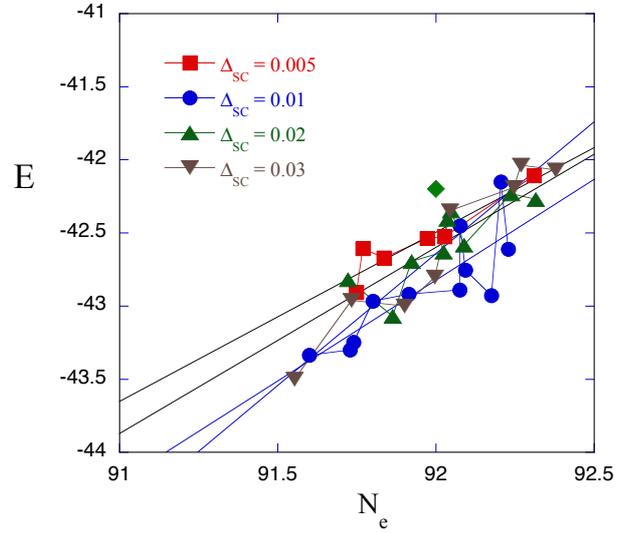}
\caption{(Color online)
The ground-state energy $E$ as a function of the electron number
$N_e$ near $N_e=92$ on a $10\times 10$ lattice where we
choose $\Delta_{SC}= 0.005$, 0.01, 0.02, and 0.03, and $\Delta_{AF}=0.20$.
The diamond indicates the ground-state energy for the AF state without the
SC order parameter.
}
\end{figure}

\begin{figure}
\centering
\includegraphics[width=8.0cm]{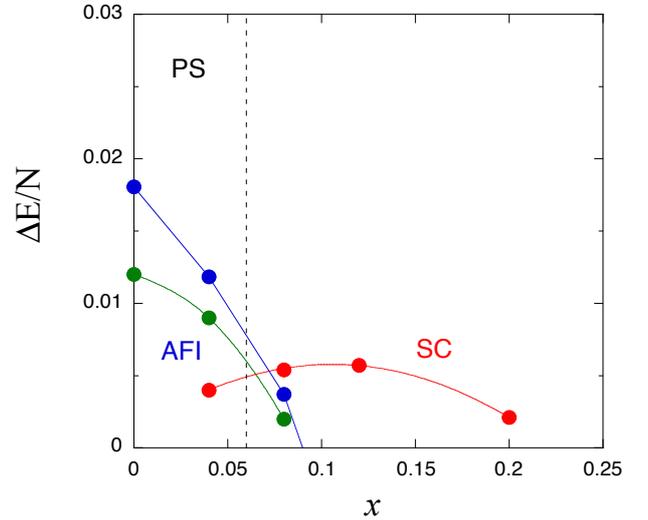}
\caption{(Color online)
The condensation energy per site as a function of the hole density
$x=1-n_e$ on a $10\times 10$ lattice, where the wave function is
$\psi_{\lambda}=\psi^{(2)}$.
We set $t'=0$ and $U/t=18$.
We used the wave function $\psi^{(4)}$ for the lower AF curve.
In the region where $x\le 0.06$, the ground state is an insulator
due to the instability toward the phase separation.
}
\end{figure}

\section{Summary}

We have investigated the ground-state properties of the 2D
Hubbard model 
by using the optimization variational Monte Carlo method.
We used the exponential-type wave function given in the
form $\exp(-\lambda S)$ with an appropriate operator $S$ and a variational
parameter $\lambda$.
With our wave function, the ground-state energy is lowered greatly
and the energy expectation value is lower than that obtained by any
other wave functions such as the Gutzwiller wave function and also
several proposed wave functions with many variational parameters.
The ground-state energy is lowered by the kinetic-energy
gain originating from $\exp(-\lambda K)$.

The AF state is very stable near half-filling
(with no carriers) in the 2D Hubbard model.
The AF correlation is suppressed as the doping rate of holes
increases.  
As the strength of the on-site Coulomb
interaction $U$ increases, a crossover occurs between weakly
correlated region and strongly correlated region.  
In the strongly correlated region, where $U$ is larger than $U_c$,
which is of the order of the bandwidth, the AF correlation is
suppressed.  A decrease in the AF correlation indicates an 
increase in spin and charge fluctuation.
This fluctuation is caused by an increase in kinetic energy and
is likely induce electron pairing.
We expect that this is an origin of high-temperature superconductivity. 

We have shown the phase diagram in the plane of $\Delta E$ (condensation
energy) and the hole doping rate $x$ for $t'=0$.  The value $t'=0$
is most favorable for superconductivity, and for nonzero $t'$ the
AF area increases.
In the underdoped region, 
where the doping rate is approximately $x < 0.09$,
AF order and superconductivity coexist.

We have also discussed the instability toward the phase separation
in the low-doping region.
The occurrence of this instability is dependent on the balance between
the kinetic energy gain of holes and the electron interaction between the adjacent
lattices such as the pairing interaction and the AF interaction. 
In the range $x\le 0.06$, the state of phase separation is realized
for $t'=0$.  In this region the AF state is an insulator, and becomes
a metal when $x$ increases. 
The phase separation area decreases as $|t'|$ increases.
The region where AF and SC phases coexist becomes larger at the same time
with increasing $-t'$.
We have also pointed out the depression in $\chi_c$ near $x=0.12$ for
negative $t'$ such as $t'= -0.1\sim -0.2$.

This work was supported by a Grant-in-Aid for Scientific
Research from the Ministry of Education, Culture, Sports, Science and
Technology of Japan (Grant No. 17K05559).
The author expresses his sincere thanks to K. Yamaji and I. Hase for useful discussions.
Part of the computations were supported by the Supercomputer
Center of the Institute for Solid State Physics, the University of
Tokyo.


\end{document}